\documentclass[preprint,preprintnumbers,amsmath,amssymb]{revtex4}

\usepackage{graphicx}
\usepackage{dcolumn}
\usepackage{bm}

\begin{document}

\title{\bf Orbital-Specific Modeling of CO Chemisorption}

\author{Sara E. Mason, Ilya Grinberg and Andrew M. Rappe}
\affiliation{  The Makineni Theoretical Laboratories, Department of
Chemistry\\ University of Pennsylvania, Philadelphia, PA 19104--6323 }%

\date{\today}

\begin{abstract}
We demonstrate that variations in molecular chemisorption energy on
different metals, different surface terminations, and different strain
conditions can be accounted for by orbital-specific changes in the
substrate electronic structure.  Our density functional theory data
set, spanning three metals, two surface terminations, and five strain
states, is fit to a single model based on tight binding.  A
crucial aspect of the model is decomposition of the $d$-band into
contributions from the five $d$ atomic orbitals.  This provides a
representation of the energy levels of the substrate that are directly
relevant to the chemisorption bond, leading to accurate prediction of
chemisorption trends.
\end{abstract}

\maketitle


Currently, great attention is focused on elucidating how surface
modification affects surface
reactivity.~\cite{Gsell98p717,Rose02p48,Schlapka03p016101,Kitchin04p156801,Abild-Pederson05p9}
Recent research shows that small changes in surface electronic
structure, induced by alloying or strain, can significantly change
surface-catalyzed reaction rates.~\cite{Bligaard04p206} A quantitative
understanding of how changes in surface geometry and electronic
structure affect surface reactivity will enable the design of more specific and more effective catalysts.  It has been shown that $\epsilon_d$, the center of
the transition metal (TM) $d$-band density of states projected on the
surface atoms (PDOS), is generally predictive of trends in
chemisorption energies ($E_{\rm chem}$) on TM
surfaces.~\cite{Hammer96p2141,Hammer97p31,Mavrikakis98p2819} However,
quantitative accuracy (model predictions accurate within 0.1~eV) is
still elusive, and for several cases there is poor or no correlation
between $\epsilon_d$ and $E_{\rm
chem}$.~\cite{Lu02p3084,Liu04p10746,Cooper05p081409R} In this paper,
using CO chemisorption as an example, we show that more rigorous
examination of the surface electronic structure coupled with a simple
modification of current chemisorption modeling enables us to achieve
this goal.

We have compiled a database of DFT molecular top site ($E_{\rm chem}$)
and dissociative bridge site ($E_{\rm dissoc}$) chemisorption energies
and electronic structure measurements for CO on Pt, Pd, and Rh (111)
and (100) surfaces.  $E_{\rm chem}$ and $E_{\rm dissoc}$ are
determined for each surface at the preferred theoretical lattice
constants as well as under in-plane strains of $\pm$1\% and $\pm$2\%,
a range easily achievable through epitaxial
mismatch.~\cite{Hrbek01p67} Studying the response of chemisorption to
strain as well as to different metals and facets deepens the study.
Strain induces relatively subtle changes in $E_{\rm chem}$ and $E_{\rm
dissoc}$ (compared to changes in metal or facet), so accurately
accounting for large and small changes is a stringent test of a
proposed theoretical model.  Since lateral stress changes inter-planar
separations, straining the systems also probes the interplay between
in-plane and inter-plane perturbations to the surface geometry.

For each metal, surface, and strain state, two values for $E_{\rm
chem}$ are determined, $E_{\rm chem}^{\rm fix}$ and $E_{\rm chem}^{\rm
rlx}$.  The former is the energy gain when the same chemisorption
geometry is fixed over the relaxed bare surface for all metals and
surfaces.  In the latter, full ionic relaxation is allowed in the top
two metal surface layers and all C and O ionic degrees of freedom.
For the dissociative systems, we only determine $E_{\rm dissoc}^{\rm
fix}$, due to the known instability of C and O atomic adsorption at
bridge sites~\cite{Ford05p159}.

We focus attention on top-site $E_{\rm chem}$ and bridge-site $E_{\rm
dissoc}$ for clarity.  The symmetries at these sites provide for zero
overlap between some $d$ orbitals and the adsorbate orbitals, making a
clear distinction between orbital-specific and orbital-averaged
models.  Analysis of the top site also facilitates separation of the
molecular chemisorption into $\sigma$ and $\pi$ contributions.
However, even at low-symmetry sites, the contributions of the five $d$
orbitals to chemisorption are unequal, making our orbital-specific
treatment more accurate in general.

DFT calculations are performed with a generalized-gradient approximation 
(GGA) exchange-correlation functional~\cite{Perdew96p3865} and
norm-conserving optimized pseudopotentials~\cite{Rappe90p1227} with
the designed nonlocal method for
metals.~\cite{Ramer99p12471,Grinberg01p201102}  Pseudopotentials
were designed using the OPIUM pseudopotential package.~\cite{Opium}
All $E_{\rm chem}$ values have been corrected using
our first-principles extrapolation procedure.~\cite{Mason04p161401R}
Metal surfaces are modeled as slabs of five layers
separated by vacuum, with the $c(4\times2)$ surface cell for (111)
surfaces and the $p(2\times2)$ surface cell for (100).
CO top site and C and O bridge site chemisorption are modeled at coverage
of $\Theta$=1/4.  Calculations are done, and values of $E_{\rm chem}$
tested to be converged within 0.02~eV, using an $8\times8\times1$ grid
of Monkhorst-Pack $k$-points, reduced by symmetry
where possible.~\cite{Monkhorst76p5188}

The PDOS for each orbital is constructed by projecting each atomic
valence pseudo-wavefunction (radial wavefunction multiplied by real
combination of spherical harmonics) of the surface atoms onto all the
Kohn-Sham orbitals.  Values of $\epsilon_d$ are then calculated as the
first moment of each PDOS.

To reduce PDOS contributions from neighboring surface atoms,
projection is performed within a sphere of radius $r_{\rm cut}$
centered about the surface atom of interest.  Standard practice is to
use a constant value for $r_{\rm cut}$ when comparing the PDOS and
associated $\epsilon_d$ values of different surfaces, and $r_{\rm
cut}=2$~a.u. is the default in some widely used DFT
packages.~\cite{Gonze02p478,ABINIT,DACAPO} However, this approach
leaves significant contributions from the orbitals of other atoms,
making the calculated value of $\epsilon_d$ dependent on $r_{\rm
cut}$, which is undesirable.

\begin{figure}[ht!]
\includegraphics[width=3.0in]{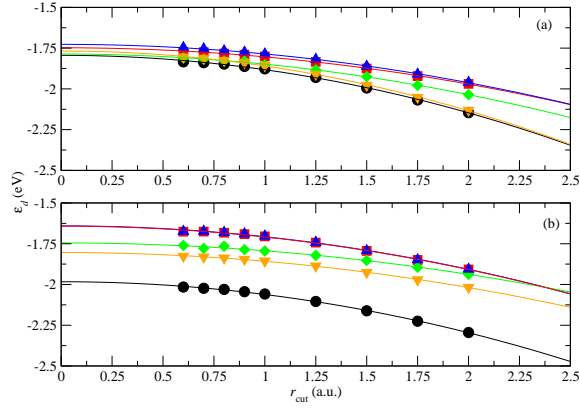}
\caption{{(a) Variation in $\epsilon_{d}$ as a function of
    projection sphere cutoff radius $r_{\rm cut}$ for Pt(111).
    $\epsilon_{x^2-y^2}$, $\epsilon_{xz}$, $\epsilon_{z^2}$,
    $\epsilon_{yz}$, and $\epsilon_{xy}$ PDOS centers are shown by
    circles, squares, diamonds, up-triangles, and down-triangles,
    respectively.  (b) Same as (a) for Pt(100).\\The graphs
    demonstrate the importance of extrapolating $r_{\rm
    cut}\rightarrow0$~a.u.: for Pt(111) the asymptotic $d$-band
    centers are more nearly equal than their large-$r_{\rm cut}$
    estimates; for Pt(100) they are more dissimilar.}}
\label{fig:Rcut}
\end{figure}

To eliminate contamination from the orbitals of
neighboring atoms, we evaluated $\epsilon_d$ at various
$r_{\rm cut}$ values. (When $r_{\rm cut}<0.5$~a.u., 
the number of FFT grid points is too small to allow
spherical sampling, so data are presented for $r_{\rm
cut}\ge0.6$~a.u.)  Figure~\ref{fig:Rcut} shows the variation of
$\epsilon_d$ with $r_{\rm cut}$ for Pt(111) and Pt(100).  It is
apparent that the asymptotic behavior of $\epsilon_d$ is not reached
until $r_{\rm cut}\ll 2$~a.u.  
Furthermore, $\epsilon_d$ values for different
surfaces, strain states and orbital angular momenta depend
differently on $r_{\rm cut}$.  To obtain accurate $\epsilon_d$ values,
we fit a purely quadratic function to the data and extrapolate
$\epsilon_d$ to $r_{\rm cut}=0$.  This procedure greatly reduces
the contribution from the orbitals of neighboring atoms, making
comparison of $\epsilon_d$ for various systems more meaningful.

\begin{figure}[ht!]
\includegraphics[width=3.3in]{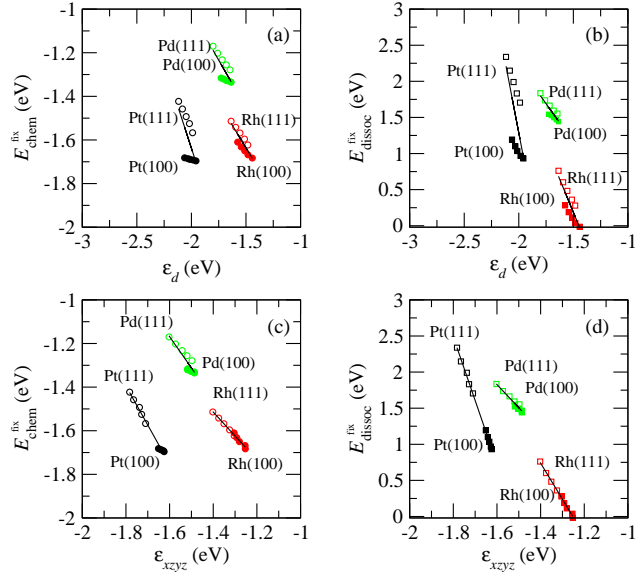}
\caption{Plots of $E_{\rm chem}^{\rm fix}$ and $E_{\rm dissoc}^{\rm fix}$
{\em vs}.\ $\epsilon_d$ and $\epsilon_{xzyz}$ for Pt (circle),
Rh (square), and Pd (diamond) (111) surfaces (open) and (100)
surfaces (filled).  Data for five lateral strain states (0\%,
$\pm$1\%, and $\pm$2\%) are shown.  Linear regressions are shown for each metal.  
(a) $E_{\rm chem}^{\rm fix}$ {\em vs}.\  $\epsilon_d$, $r_{\rm cut}$=2~a.u. 
(b) $E_{\rm dissoc}^{\rm fix}$ {\em vs}.\  $\epsilon_d$, $r_{\rm cut}$=2~a.u. 
(c) $E_{\rm chem}^{\rm fix}$ {\em vs}.\ 
$\epsilon_{xzyz}$, $r_{\rm cut}\rightarrow$0~a.u. 
(d) $E_{\rm dissoc}^{\rm fix}$ {\em vs}.\  $\epsilon_d$
and $\epsilon_{xzyz}$, $r_{\rm cut}\rightarrow$0~a.u.}
\label{fig:AssocDissoc}
\end{figure}

While our data confirm that $E_{\rm chem}$ qualitatively tracks with
$\epsilon_d$ for the (111) surfaces, they reveal shortcomings of using
$\epsilon_d$ for modeling bonding on different facets.  From
Figure~\ref{fig:AssocDissoc}, we see that $E_{\rm chem}$ values on
different facets differ by 0.06--0.25~eV (0.15--0.85~eV for $E_{\rm
dissoc}$) even though they have the same $\epsilon_d$ and the same
metal.  Figure~\ref{fig:AssocDissoc} and
Table~\ref{table:EchemRESULTS} show this is still true for
$\epsilon_d$ with $r_{\rm cut}\rightarrow0$~a.u.  Therefore, neither
$E_{\rm chem}^{\rm fix}$ nor $E_{\rm dissoc}^{\rm fix}$ can be fit as
a single function of $\epsilon_d$, when both (111) and (100) facets
are considered.

By contrast, we find that facet dependence of
the chemisorption energies can be fit as a
single linear function of $\epsilon_{xzyz}$, the band center 
of the $d_{xz}$ and $d_{yz}$ orbitals.  Figure~\ref{fig:AssocDissoc}
shows that single linear regressions of data for both facets are
accurate to within 0.05~eV in all cases.  This result demonstrates
that focusing on the metal orbitals involved in bonding simplifies
the observed chemisorption behavior and enables robust modeling.

Similarly, Table~\ref{table:EchemRESULTS} shows that the response of
$E_{\rm chem}$ and $E_{\rm dissoc}$ to strain is reflected in
$\epsilon_{xzyz}$ but not $\epsilon_d$.  The tunability of $E_{\rm
  chem}$ and $E_{\rm dissoc}$ through strain is two to ten times
greater on the (111) surfaces.  However, $d\epsilon_d/ds$ is identical
within computational precision for the two facets of each metal, so
$d\epsilon_d/ds$ is uncorrelated with $dE_{\rm chem}/ds$ and $dE_{\rm
  dissoc}/ds$.  The chemisorption tunability trend is strongly
reflected in $d\epsilon_{xzyz}/ds$, suggesting that the response of
the $d_{xz}$ and $d_{yz}$ orbitals plays a key role in modeling
top-site molecular and bridge-site dissociative adsorption of CO.

\begin{table}[ht!]
\caption{$E_{\rm chem}^{\rm fix}$, $E_{\rm chem}^{\rm
rlx}$, $E_{\rm dissoc}^{\rm fix}$, $\epsilon_d$, and
$\epsilon_{xzyz}$ in eV
for the (111) and (100) surfaces of Pt, Rh,
and Pd.  
The slope of each quantity with respect
to in-plane strain, ($\frac{d}{ds}$), is also reported, 
in units of meV/\%strain.  
($\epsilon_d$ and $\epsilon_{xzyz}$ use $r_{\rm cut}\rightarrow0$~a.u.)}
\begin{tabular}{llr|lr|lr|lr|lr}
              & \multicolumn{2}{c}{$E_{\rm chem}^{\rm fix}$ $\frac{dE}{ds}$}&  \multicolumn{2}{c}{$E_{\rm chem}^{\rm rlx}$  $\frac{dE}{ds}$}& \multicolumn{2}{c}{$E_{\rm dissoc}^{\rm fix}$ $\frac{dE}{ds}$}&  $\epsilon_d$ &$\frac{d\epsilon}{ds}$& $\epsilon_{xzyz}$&$\frac{d\epsilon}{ds}$\\
\hline
Pt(111)     &   -1.49 &-36 & -1.58 &-46 & 1.99 &-158  & -1.77 & 20& -1.74 &18 \\
Pt(100)     &   -1.69 & -3 & -1.81 & -5 & 1.03 & -65  & -1.76 & 20& -1.63 & 7 \\
Rh(111)     &   -1.56 &-27 & -1.68 &-24 & 0.48 &-120  & -1.38 & 31& -1.35 &25 \\
Rh(100)     &   -1.65 &-18 & -1.74 & -9 & 0.11 & -74  & -1.36 & 29& -1.28 &14 \\
Pd(111)     &   -1.23 &-25 & -1.26 &-29 & 1.66 & -70  & -1.58 & 32& -1.54 &26 \\
Pd(100)     &   -1.32 &-5 & -1.36 & -6 & 1.50 & -29  & -1.56 & 19& -1.50 & 9 \\
\end{tabular}
\label{table:EchemRESULTS}
\end{table}

To explain the observed trends in $E_{\rm chem}$ as the metal
identity, facet, and strain state are varied, we reconsider the model
for CO chemisorption.  A second-order perturbative picture of
chemisorption involving interacting molecular and metal orbitals is
simple and intuitive, and has been known for some
time.~\cite{Hoffmann88p601,Blyholder64p2772} Hammer, Morikawa, and
N{\o}rskov (HMN) achieved significant success modeling molecular
interactions with solid surfaces with a single perturbative term
involving $\epsilon_d$, the $d$-band center.~\cite{Hammer96p2141}
Guided by the trends presented above, we cast an orbital-specific
analysis (OS) in the HMN model form, modeling top-site chemisorption as a
perturbative interaction between molecular orbitals and the $d$-band
PDOS of each spatial orbital:

\begin{eqnarray}
E_{\rm chem}^{\rm OS} = E_{sp} 
-4\left\{ \frac{fV_{\pi}^{2}}{\epsilon_{2\pi^*}-\epsilon_{xzyz}}+fS_{\pi}V_{\pi}\right\} \nonumber\\-2\left\{ \frac{(1-f)V_{\sigma}^{2}}{\epsilon_{z^{2}}-\epsilon_{5\sigma}}+(1+f)S_{\sigma}V_{\sigma}\right\}
\label{eqn:specific}
\end{eqnarray}
\noindent where $f$ is the idealized filling of the metal $d$ bands,
$V$ and $S$ are perturbation matrix elements and overlap integrals,
respectively, labeled by symmetry, $\epsilon_{2\pi^*}$ and
$\epsilon_{5\sigma}$ are the CO molecular orbital energies,
$\epsilon_{xzyz}$, the band center of the $d_{xz}$ and $d_{yz}$
orbitals and $\epsilon_{z^{2}}$ is the band center of the $d_{z^{2}}$
orbitals.~\footnote{Taking the top site as an example, $S_{\pi,i}\ne0$
for $d_{xz}$ and $d_{yz}$ only, and $S_{\sigma,i}\ne0$ for $d_{z^2}$.
HMN values of $\epsilon_{2\pi^*}$=2.5~eV and $\epsilon_{5\sigma}$=7~eV
are used.  Moderate changes to these values re-scales the dependence of
$E_{\rm chem}$ on $\epsilon_{d}$ without significantly changing any
presented results.}
As in the original HMN model, $\alpha$ and $\beta$ are introduced as
fitting parameters common to all metals, and $V_{\pi}^{2} \approx \beta
V_{sd}^{2}$ and $S_{\pi} \approx -\alpha V_{\pi}$.  From our analysis
of DFT orbitals, we find that $S_{\sigma}$/$S_{\pi}$ to be sensitive
in the limit of desired accuracy to both metal identity and adsorption
geometry.  Our overlap analysis gives $S_{\sigma}$/$S_{\pi}$ is 1.182,
1.156, and 1.200 for Pt, Rh, and Pd, respectively.  $E_{sp}$ is found
to be -0.15~eV from DFT calculations on Al
surfaces~\cite{Mason06preprint} and assumed to be independent of metal
identity, facet, or strain.  For the (111) and (100) surfaces, we find
that $f$ for each of the decomposed $d$-bands is well approximated by
the idealized filling of the metal $d$-bands, $f = (\nu -1)$/10, where
$\nu$ is the valence of the metal atom.

The corresponding conventional orbital-averaged (OA) model form is given by:
\begin{eqnarray}
E_{\rm chem}^{\rm OA} = E_{sp} 
-4 \left\{ \frac{fV_{\pi}^{2}}{\epsilon_{2\pi^*}-\epsilon_{d}}+fS_{\pi}V_{\pi}\right\} \nonumber\\-2\left\{ \frac{(1-f)V_{\sigma}^{2}}{\epsilon_{d}-\epsilon_{5\sigma}}+(1+f)S_{\sigma}V_{\sigma}\right\}
\label{eqn:averaged}
\end{eqnarray}

\begin{figure}
\includegraphics[width=3.3in]{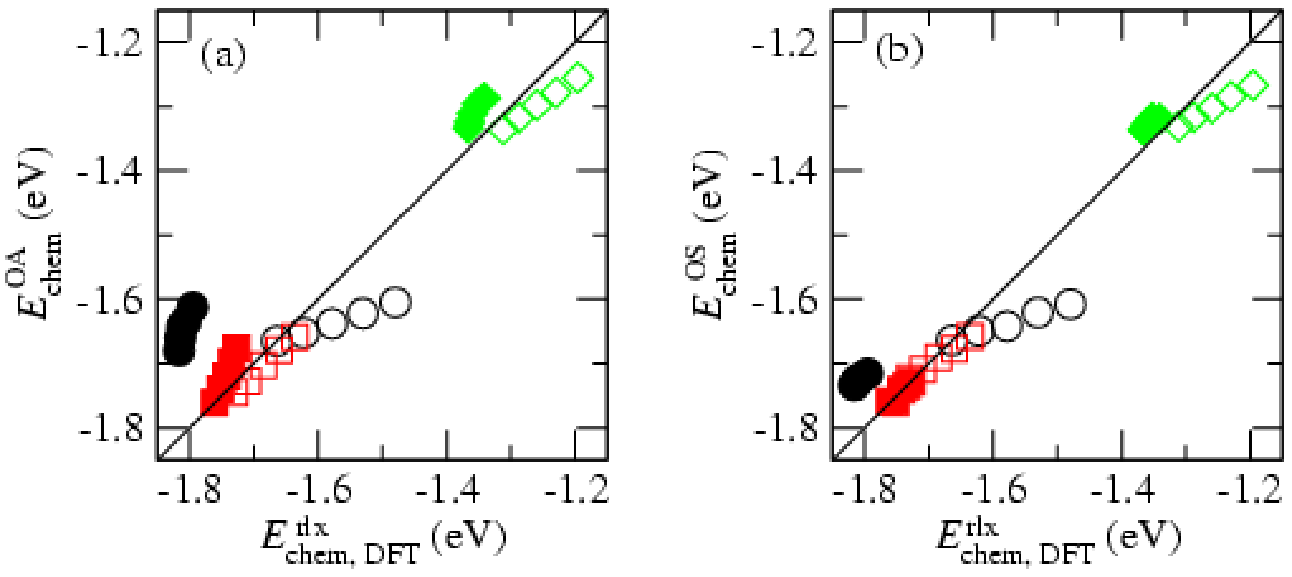}
\caption{{Correlation plots of modeled $E_{\rm chem}$ and $E_{\rm
chem, DFT}$.  Pt (circle), Rh (square), and Pd (diamond) (111) surfaces
(open) and (100) surfaces (filled).  Data for five lateral strain
states (0\%, $\pm$1\%, and $\pm$2\%) are shown.  
(a)$E_{\rm chem}^{\rm OA}$ (Equation~\ref{eqn:averaged}) {\em vs}. $E_{\rm chem, DFT}^{rlx}$. 
(b)$E_{\rm chem}^{\rm OS}$ (Equation~\ref{eqn:specific}) {\em vs}. $E_{\rm chem, DFT}^{rlx}$.\\The plots show that orbital-specific modeling (Equation~\ref{eqn:specific}) is required to achieve the same quality of correlation for both (111) and (100) facets. 
}}
\label{fig:Correlations}
\end{figure}

We fit the data for both fix and rlx chemisorption systems to the Equation~\ref{eqn:specific} and
Equation~\ref{eqn:averaged}.~\footnote{For the fix data, the best fit
to Equation~\ref{eqn:specific} is achieved when $\alpha$=0.0619~eV and
$\beta$=1.052~eV$^{2}$, and the best fit to
Equation~\ref{eqn:averaged} is achieved when $\alpha$=0.0578~eV and
$\beta$=0.938~eV$^{2}$.  For the rlx data, the best fit to
Equation~\ref{eqn:specific} is achieved when $\alpha$=0.0619~eV and
$\beta$=1.050~eV$^{2}$, and the best fit to
Equation~\ref{eqn:averaged} is achieved when $\alpha$=0.0604~eV and
$\beta$=1.048~eV$^{2}$.}  Figure~\ref{fig:Correlations} shows the
correlation between DFT and model values for the more realistic rlx
chemisorption systems.  We calculate the root-mean-square error (RMSE)
as an evaluation of the model, considering the (111) and (100) data
separately and combined.  When Equation~\ref{eqn:averaged} is used to
fit the data, the RMSE is 0.051 and 0.100~eV for the (111) and (100)
surfaces respectively, and 0.079~eV overall.  When
Equation~\ref{eqn:specific} is used to fit the data, the RMSE is 0.052
and 0.051~eV for the (111) and (100) surfaces respectively, and
0.052~eV overall.  This shows that the more sophisticated model form
of Equation~\ref{eqn:specific} is required to achieve the same level
of accuracy for these two surfaces.

We now address why different levels of model sophistication are needed
to achieve the same accuracy in predicted $E_{\rm chem}$ on the (111)
and (100) surfaces.  First, we consider the salient electronic
structure differences between the (111) and (100) surface facets.  The
$dd$ metal bonding can be decomposed by symmetry into $\sigma$, $\pi$,
and $\delta$ contributions.~\cite{Sutton93} The square lattice of the
(100) surface allows for strong $dd\sigma$ overlap between neighboring
$d_{x^2-y^2}$ orbitals.  Our DFT data show that $\epsilon_{x^2-y^2}$
is significantly lower on (100) surfaces than on (111), as shown for
Pt in Figure~\ref{fig:Rcut}.  Since (100) surface atoms have eight
nearest neighbors while the (111) atoms have nine, the other $d$
orbitals are less stable on (100) surfaces to varying extents, with
$\epsilon_{xz}$ and $\epsilon_{yz}$ significantly higher on (100) than
(111).  Averaging all the $d$ orbitals causes the rise in
bonding-relevant $\epsilon_{xz}$ and $\epsilon_{yz}$ to be masked by
the drop in bonding-irrelevant $\epsilon_{x^2-y^2}$, so that even
though $\epsilon_{xzyz}$ closely tracks the increase in $E_{\rm chem}$
on the (100) surface relative to the (111) surface, the averaged
$\epsilon_d$ does not.

When tensile strain is applied to a TM surface, the weakened in-plane
bonding destabilizes the $d$ orbitals.  $\epsilon_d$ shifts upward,
leading to stronger $E_{\rm chem}$.~\cite{Hammer96p2141} This basic
prediction was confirmed in the DFT study of Mavrikakis {\em et
al}.~\cite{Mavrikakis98p2819} However, different $d$ orbitals shift by
different amounts, based on their orientations relative to the
surface.

In addition, the inter-planar spacing between the top two metal layers
($r_{12}$) responds to the strain, and this further affects substrate
electronic structure, again in an orbital-specific way.  Tensile
lateral strain usually decreases $r_{12}$, while compression increases
$r_{12}$.  The bonding-relevant $d_{xz}$ and $d_{yz}$ orbitals of the
top layer have the strongest interaction with the second layer atoms,
so the relaxation of $r_{12}$ significantly reduces the effect of
lateral strain for these orbitals.  This is why $d\epsilon_{xzyz}/ds$
is less than $d\epsilon_d/ds$ for all surfaces studied
(Table~\ref{table:EchemRESULTS}).

The effect of $r_{12}$ relaxation on strain tunability is also
strongly facet-dependent.  On the more open (100) surface, relaxations
of $r_{12}$ are larger, making $d\epsilon_{xzyz}$ smaller for each
(100) facet studied than for the corresponding (111) facet.  This
explains why the tunability of $E_{\rm chem}$ (fix and rlx) and
$E_{\rm dissoc}^{\rm fix}$ are much lower on (100) surfaces than on
(111) surfaces.

Our demonstration that a single orbital-specific chemisorption model
can be applied to different facets, strains, and metals, has
implications for the modeling and design of more realistic catalyst
surfaces.  DFT studies have found that reactions on late TMs are more likely to proceed on defects such as steps and
kinks.~\cite{Liu03p1958} The model presented suggests that one should
examine how different nearest-neighbor and inter-planar separations
affect the orbital-specific electronic structure, and predict
chemisorption properties accordingly.

Incorporating the effects of strain and $r_{12}$
relaxation on the relevant PDOS centers greatly improved chemisorption
modeling.  We therefore suggest that if further couplings between PDOS
centers and adsorbate structure can be parameterized, the resulting
model could offer even greater accuracy and broader applicability.

In conclusion, we use the energy levels of the substrate $d$-band
projected onto the substrate atomic orbitals and their overlap with
the CO bonding molecular orbitals, in a second-order perturbation
theory-type model for chemisorption.  The resulting model is able to
account for changes in $E_{\rm chem}$ on different surface facets
under different conditions of strain.  We have also shown that trends
in the dissociative chemisorption of CO at bridge site are governed by
the same orbital-specific factors.  The results shown here should be
generally valid for other molecular and atomic adsorption on higher
index surfaces and for perturbations other than strain.

This work was supported by the Air Force Office of Scientific
Research, Air Force Materiel Command, USAF, under grant number
FA9550--04--1--0077, and the NSF MRSEC Program, under Grant
DMR05--20020.  Computational support was provided by the Defense
University Research Instrumentation Program, and the High Performance
Computing Modernization Office of the Department of Defense.  SEM
acknowledges Valentino R. Cooper for fruitful exchange of ideas.

\bibliography{thebibliography}

\clearpage

\clearpage

\clearpage

\end{document}